# Quantitative discrimination between oil and water in drilled bore cores via fast-neutron resonance transmission radiography[+]

D. Vartsky[a,*], M. B. Goldberg[b], V. Dangendorf[c], I. Israelashvili[a,e], I. Mor[d], D. Bar[d], K. Tittelmeier[c], M. Weierganz[c], A. Breskin[a]

[a] Weizmann Institute of Science, Rehovot, 76100, Israel
[b] Herzbergstr. 20, 63584 Gründau, Germany
[c] Physikalisch-Technische Bundesanstalt (PTB), 38116 Braunschweig, Germany
[d] Soreq NRC, Yavne 81800, Israel
[e] Nuclear Research Center of the Negev, P.O.Box 9001, Beer Sheva, Israel

**Abstract**

A novel method based on Fast Neutron Resonance Transmission Radiography is proposed for non-destructive, quantitative determination of the weight percentages of oil and water in cores taken from subterranean or underwater geological formations. The ability of the method to distinguish water from oil stems from the unambiguously-specific energy dependence of the neutron cross-sections for the principal elemental constituents. Monte-Carlo simulations and initial results of experimental investigations indicate that the technique may provide a rapid, accurate and non-destructive method for quantitative evaluation of core fluids in thick intact cores, including those of tight shales for which the use of conventional core analytical approaches appears to be questionable.

**Keywords:** Petrophysical analysis, core analysis, core fluid content, Fast Neutron Resonance Radiography,

---

[*] Corresponding author. Tel.: + 972.50.6292126; fax: + 972. 8.9342611
*E-mail address:* david.vartsky@weizmann.ac.il
[+] Patent pending

## 1. Introduction

When a whole drill core of reservoir rock extracted from oil or gas well enters a modern petro-physical laboratory, it may undergo several non-destructive tests, such as a gamma-ray scan which correlates core depth with log depth, or a CT scan which can provide a 3-D image of a the whole core, providing information on its internal features and fractures.
The core is then cut into slabs, disks or plugs that subsequently undergo a procedure of cleaning and fluid extraction, to obtain samples of dry rock and the associated quantities of extracted oil and water (Andersen et al., 2013). These samples serve for determining the critical core properties required to decide whether the well in question can be economically exploited. The parameters relevant to such decision are: **porosity**-storage capacity for reservoir fluids, **permeability**-reservoir flow capacity, as well as **fluid saturation**-fluid type and content (American Petroleum Institute, 1998, Ubani et al., 2012).

Nuclear geophysics is a discipline that assists oil, gas and uranium exploration, both in nuclear borehole logging and analysis of core samples (Borsaru M., 2005). Middleton et al, 2001, investigated thermal neutron radiography to estimate rock's porosity and relative fluid saturation in 5 mm thick rock slices. The use of thermal neutrons does not permit distinguishing between water and oil, because it relies solely on the attenuation of hydrogen. De Beer et al, 2004 also used thermal-neutron radiography to provide internal structure images of rocks, in order to determine the effective porosity of the object. Nshimirimana et al, 2014 examined the precision of porosity calculations in 14-17 mm thick rock samples using thermal neutron radiography.

Lanza et al, 1991 investigated thermal neutron computerized tomography to image the distribution of hydrogenous liquids (oil or water) in a 25.4 mm diameter core. As in the above-mentioned studies it cannot distinguish between oil and water either. In certain cases deuterated water is introduced into the porous media, in order to study immiscible fluid flow by thermal neutron tomography (Murison et al. 2015). A recent review (Perfect et al, 2014) of thermal-neutron imaging of hydrogen-rich fluids in geo-materials discusses the non-destructive visualization of such fluids within diverse porous media.

In this paper we present a new method based on Fast Neutron Resonance Transmission (FNRT) radiography that can determine non-destructively and quantitatively the content of different fluids in the core.

## 2. Fast Neutron Resonance Transmission Radiography

FNRT is a method that exploits characteristic structures (resonances) in the neutron attenuation of the analysed object constituents to determine the identity and proportions of substances within it. A typical neutron energy range is 1-10 MeV.

In FNRT the inspected object is irradiated with a broad spectrum of neutrons in the above-mentioned energy range. Depending on the nature of the inspected object the transmitted neutron spectrum will exhibit dips and peaks at specific energies.

Therefore, the transmitted neutron spectrum carries information about the composition of the object.

FNRT has been applied in the past for detecting low-Z (light) elements, such as H,C, N, and O, in order to determine composition of agricultural products and detect contraband (Overley, 1985). A system for detection of explosives in air-passenger bags based on this method has also been constructed and tested. (Overley et al, 1997; Miller et al, 1997). All the above investigators used accelerator-based, nanosecond-pulsed, broad-energy neutron beams for interrogating the objects of interest.

High position resolution fast neutron detectors required for FNRT have been developed in the last few years (Dangendorf et al, 2008, Mor et al, 2009, Brandis et al., 2012, Israelashvili et al. 2015). For a review on the prospects of FNRT and its requirements for instrumentation, see Vartsky, 2006.

Fig. 1 shows the energy dependence of the mass attenuation coefficients of silica (the principal constituent of sandstone core), oil and water taken from Evaluated Nuclear Data File (ENDF, 2015). It can observed that the attenuation coefficients of the three substances exhibit different characteristic behaviour with neutron energy due to resonances of the most abundant elements in materials, such as carbon in oil, oxygen in water and oxygen and silicon in silica. Thus the signature of each material is unique.

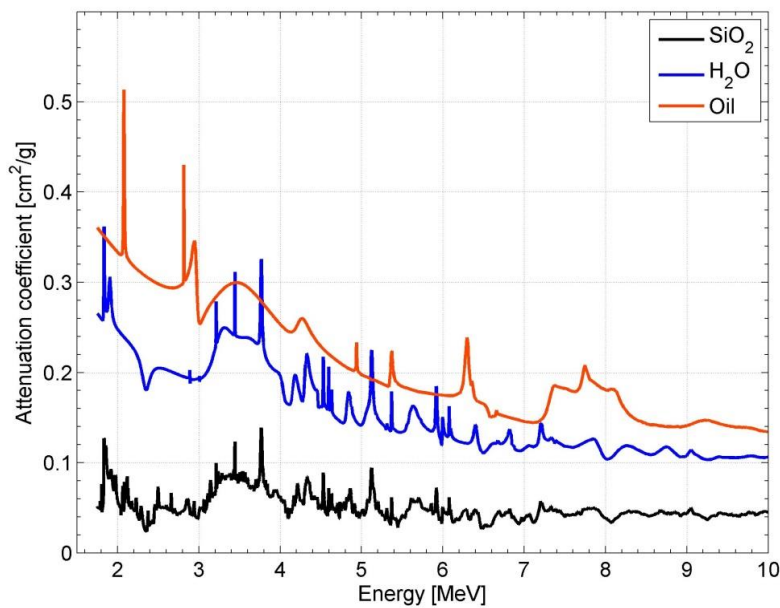

Fig. 1 Mass attenuation coefficients of silica, oil and water vs. neutron energy

Fig. 2 schematically shows the FNRT irradiation configuration. The object is subjected to a neutron beam of a broad spectral distribution in the energy range 2-10 MeV. The transmitted neutron spectrum is detected by a fast-neutron position-sensitive detector to provide high-resolution imaging capability.

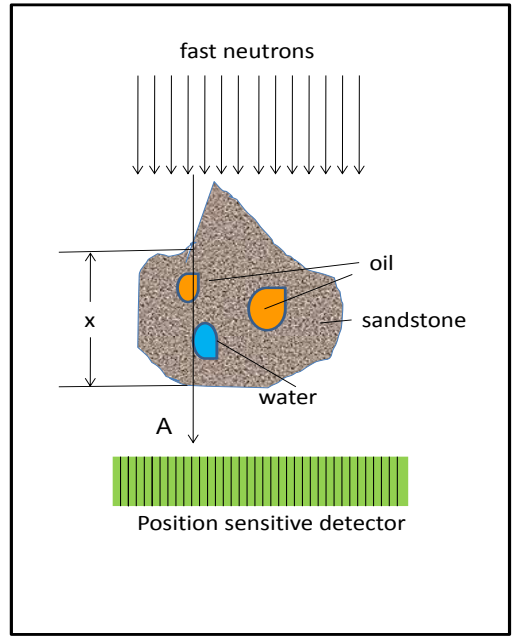

Fig. 2 Schematic description of FNRT irradiation configuration of a rock sample. The arrow marked "A" stands for a beam of fast neutrons that traverses the thickness x of the sample and impinges on a specific pixel within the array.

If we assume that the inspected object, such as an oil-drilling core, consists mainly of porous rock (sandstone, limestone), oil and water, the ratio $R_i$ of the transmitted-to-incident neutron flux at an energy **i** and at the position indicated in the drawing by the long arrow (A) is:

$$R_i = \exp[-\boldsymbol{\mu}_i^s \rho_s x + \boldsymbol{\mu}_i^o \rho_o x + \boldsymbol{\mu}_i^w \rho_w x)] \quad \text{Eq.1}$$

Where $\boldsymbol{\mu}_i^s$, $\boldsymbol{\mu}_i^o$, $\boldsymbol{\mu}_i^w$ and $\rho_s x$, $\rho_o x$, $\rho_w x$ are the mass attenuation coefficients and areal densities of sandstone, oil, and water, respectively (see also Fig. 1).

Eq. 1 strictly holds only for a thin pencil beam or if the composition of the sample in the lateral direction (normal to the beam) is uniform within the area seen by the detector pixel. If a lateral non-uniformity in sample composition occurs within the pixel area, the result will be biased. This is a common problem in transmission radiometry and several methods for correcting it have been devised (Hampel and Wagner, 2011).

Analogously, the densities ρ in Eq. 1 are not the intrinsic densities of the substances, they represent the mean densities averaged over the trajectory x.

Since the spectrum may consist of **n** discrete neutron-energies, one can write n such equations. By taking a natural logarithm of $R_i$ one obtains a set of n linear equations where the areal densities are the unknowns of interest. This is an over-determined system, in which there are n linear equations with 3 unknowns (of these **n**, not all have the same sensitivity: in other words, depending on the element in question, the effective **n** may be considerably smaller than the nominal **n**. Such a problem can be solved by a least squares solution or Bayesian minimization (Lunn et al, 2003).

Once a solution for the 3 areal densities is found in a given detector pixel, we can determine the areal-density-ratio of oil or water to that of the dry rock. This yields the weight fraction of oil or water in the traversed core, independent of sample thickness or shape. One can now display the map of ratios for each individual pixel. Alternatively, by multiplying each pixel areal density by a pixel area we obtain the mass of each component in a volume defined by pixel area and height x and by integrating over all pixels obtain the total weight of oil, water and dry rock in the entire core, from which the average weight fractions of oil and water in the core can be determined.

The values of the mass attenuation coefficients vs. neutron energy must be determined experimentally for a given system using calibrated standards of pure rock, oil and water. This calibration procedure is necessary since there could be significant differences between rock and oil types from one drilling site to another. If such standards are unavailable, one may use calibrated elemental standards, such as Si, O, C, H, Ca and Mg to measure their mass attenuation coefficients. In such a case, solving Eq. 1 will yield elemental areal densities, from which one can deduce the content of oil and water in the core.

Provided the position resolution of the detector is fine enough, such that the lateral composition variations within the pixel area can be neglected, the method measures the average liquid/dry-core weight-ratio in the path traversed by the fast neutrons regardless of the object shape, thickness or distribution. As the method is fast (scans of few minutes/view) it should permit screening of appreciable lengths of a core within relatively short times.

It must be noted that the fluid weight fractions in the sample are determined independently, thus the oil-to-rock weight-ratio is independent of water content. This is not the case in conventional thermal neutron radiography, where the assigned water and oil contents are unavoidably correlated due to the lack of elemental specificity (the dominant attenuation by hydrogen is common to both).

3. **Monte-Carlo calculations**

In order to validate the method, we performed a series of GEANT 4 Monte-Carlo (Agostinelli et al, 2003) calculations for several configurations of core materials containing various quantities of oil, water and a mixture of oil and water. An incident thin pencil neutron beam with uniformly distributed energy spectrum ranging from 2-10 MeV, impinged on a core sample. Various quantities of uniformly distributed oil, water and a combination of oil and water were added to the sample. The core materials were dry sandstone (DS-bulk density 2.3 g/cc, 10 cm thick) and sand (S-bulk density 1.55 g/cc, 7.8 cm thick). The latter material was simulated in order to compare to experimental results which were carried out with sand and are described in section 4.

In order to test the influence of counting statistics on the reconstructed values, the Monte-Carlo simulation of the 10 cm thick sandstone sample was performed for 3 different numbers of incident neutrons: N= $10^7$, $10^6$ and $10^5$ neutrons.

Simulations of neutron transmission through pure standard materials (DS, S, oil and water), necessary for determination of their mass attenuation coefficients, were performed with $10^8$ incident neutrons.

The attenuated neutron spectra were normalized to a flat spectrum determined separately (using $2 \cdot 10^8$ incident neutrons), without the absorbing object in the beam, resulting in transmission spectra.

3.1 Analysis of 10 cm thick sandstone sample

Fig. 3 shows the transmission neutron spectra for the 10 cm thick DS core containing various percentages of oil or water. The total number of incident neutrons (integrated over all neutron energies) impinging on the sample was $1 \times 10^7$. The detected number of neutrons ranged from $1.3 \cdot 10^6$ to $3 \cdot 10^6$ counts. As can be observed, the spectrum is dominated by the shape of the DS spectrum; nevertheless, the proportions of the various peaks are different for each configuration.

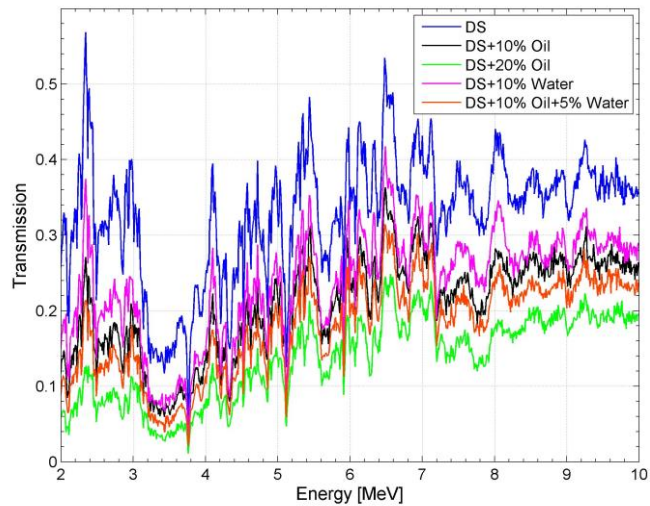

Fig. 3 Calculated transmission spectra through a 10 cm thick sample containing: sandstone (DS), DS+10% oil, DS+20% oil, DS+10% water and DS+10% oil+5% water. Total number of neutrons impinging on sample N=$10^7$. (quotes of oil and water in percent weight)

Reconstruction of sample composition followed the procedure of Mor et al, 2015 employing a Bayesian analysis. The analysis was carried out using the software WinBUGS (Lunn et al., 2000), an interactive Windows version of the BUGS program (Bayesian inference Using Gibbs Sampling) developed by the Medical Research Center (MRC) and Imperial College of Science, Technology and Medicine, UK.

The Bayesian analysis provides a probability distribution of the areal density for each constituent, indicating whether it is likely to be found in the inspected material and what is the most probable areal density. Substances which are likely to be present in the inspected sample have probability distributions that are Gaussian in shape and exhibit standard deviations within a few percent of the mean, whereas the frequency distributions of areal density for substances which are not likely to be present in the inspected sample, have probability densities that peak at or around 0 and are skewed to the extent that the standard deviation is of the order of 30 % of the mean or higher.

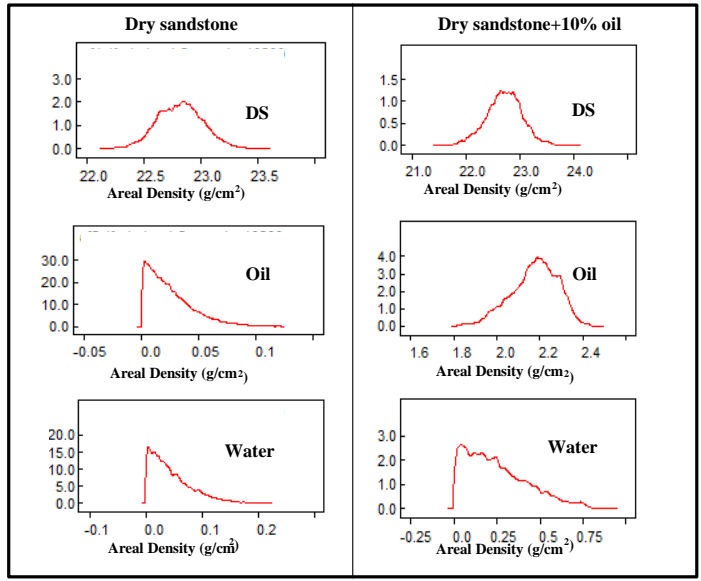

Fig 4. Areal density frequency distributions of DS, oil and water in: **left)** dry sandstone, **right)** sandstone containing 10% oil by weight. N=$10^7$ neutrons

Fig. 4 shows the resulting Bayesian reconstruction of simulated areal density distributions of dry sandstone, oil and water for two configurations. The distributions on the left are for a dry sandstone core and those on the right for a core containing 10% oil by weight. As evident from the distributions, the sample on the left does not contain any liquid, while the distributions on the right indicate that a significant amount of oil (standard deviation: 3.5%) is present and suggest a small likelihood for water-(standard deviation: 38%) as well. Table 1 provides a summary of reconstructed areal densities and their standard deviations for all sandstone core configurations.

**Table 1:** Summary of reconstructed areal densities of sandstone core configurations

N=$10^7$ neutrons per simulation

| Configuration | DS [g/cm$^2$] | Oil [g/cm$^2$] | Water [g/cm$^2$] |
|---|---|---|---|
| **Pure DS** | 22.99±0.06 (23*) | 0.01±0.01 (0*) | 0.02±0.07 (0*) |
| **DS+10% oil** | 22.83±0.12 (23*) | 2.22±0.04 (2.3*) | 0.15±0.28 (0*) |
| **DS+20% oil** | 22.97±0.12 (23*) | 4.58±0.04 (4.6*) | 0.07±0.05 (0*) |
| **DS+10% water** | 23.20±0.11 (23*) | 0.07±0.05 (0*) | 2.16±0.07 (2.3*) |
| **DS+(10% oil+5%H$_2$O)** | 23.05±0.13 (23*) | 2.3±0.05 (2.3*) | 1.12±0.08 (1.15*) |

*Expected value

The ratio of reconstructed areal density of oil (or water) to that of DS yields the weight percentage of each fluid in the core. Fig 5 shows the percentages by weight of oil and water in each configuration.

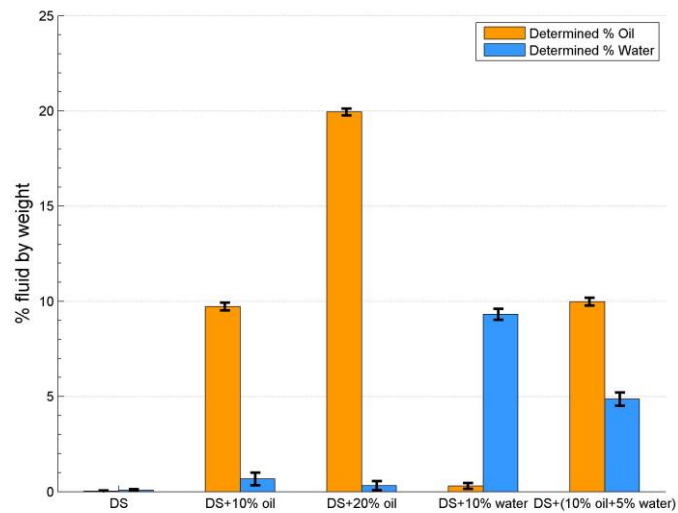

Fig. 5 Weight percentages of oil (orange) and water (blue) in 10 cm thick sandstone samples for dry sandstone (DS), DS+10% oil, DS+20% oil, DS+10% water and DS+(10% oil+5% water)

The influence of counting statistics on the reconstructed values of fluid content is shown in Table 2.

Table 2 Reconstructed % fluid in sandstone vs. number of incident neutrons per simulation.

| No. of neutrons → | $10^7$ neutrons | | $10^6$ neutrons | | $10^5$ neutrons | |
|---|---|---|---|---|---|---|
| Configuration ↓ | % oil | % water | % oil | % water | % oil | % water |
|  |  |  |  |  |  |  |
| **Dry Sandstone (DS)** | 0.03±0.03 | 0.07±0.05 | 0.10±0.08 | 0.16±0.14 | 0.30±0.25 | 0.65±0.5 |
| **DS+10% oil** | 9.72±0.20 | 0.67±0.0.33 | 9.32±0.55 | 1.41±0.88 | 7.52±1.5 | 4.8±2.7 |
| **DS +20% oil** | 19.94±0.18 | 0.32±0.24 | 20.03±0.48 | 0.55±0.55 | 19.1±1.5 | 1.7±1.7 |
| **DS+10% $H_2O$** | 0.3±0.16 | 9.31±0.29 | 0.51±0.35 | 9.34±0.67 | 2.02 ±1.2 | 6.8±2.3 |
| **DS+(10% oil+5% $H_2O$)** | 9.98±0.20 | 4.86±0.35 | 10.02±0.59 | 5.62±1.05 | 8.69±1.4 | 3.7±2.3 |

The absence of fluids in dry sandstone is always determined with high accuracy. However, for samples containing fluids, both the accuracy and precision of the reconstructed fluid content deteriorate when the total number of incident neutrons is reduced below $10^6$. Based on this study, it is not advisable to examine such a sample with less than $10^6$ incident neutrons integrated over the entire spectrum.

3.2 Analysis of a 7.8 cm thick sand sample

The next simulated cases were the experimental configurations (described in the following section) of 7.8 cm thick boxes filled with: dry sand (bulk density 1.55 g/cc) or mixtures of dry sand containing 10% or 20% oil and 10% water by weight. The total number of incident neutrons for each case was $10^7$. Table 3 summarizes the results obtained in this simulation and Fig 6 shows the percentages by weight of oil and water in each configuration.

**Table 3:** Summary of reconstructed areal densities of 7.8 cm sand samples.

$N=10^7$ neutrons per simulation

| Configuration | S [g/cm$^2$] | Oil [g/cm$^2$] | Water [g/cm$^2$] |
|---|---|---|---|
| **Pure Sand (S)** | 12.09±0.04 (12.09*) | 0.006±0.005 (0*) | 0.009±0.008 (0*) |
| **S+10% oil** | 12.00±0.05 (12.09*) | 1.21±0.02 (1.21*) | 0.03±0.03 (0*) |
| **S+20% oil** | 12.22±0.05 (12.09*) | 2.41±0.02 (2.42*) | 0.02±0.02 (0*) |
| **S+10% water** | 12.18±0.06 (12.09*) | 0.036±0.023 (0*) | 1.15±0.04 (1.21*) |
| **S+(10% oil/5%H$_2$O)** | 12.09±0.07 (12.09*) | 1.22±0.03 (1.21*) | 0.6±0.05 (0.605*) |

*Expected value

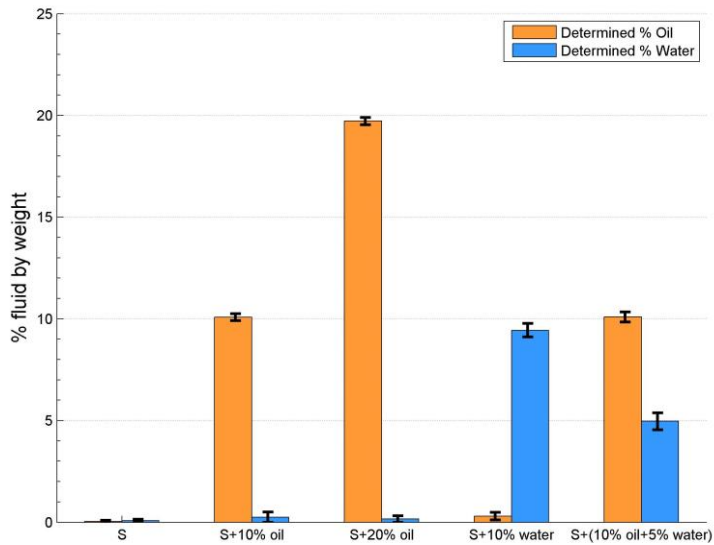

Fig. 6 Weight percentage of oil (orange) and water (blue) in 7.8 cm thick sand sample for dry sand (S), S+10% oil, S+20% oil, S+10% water and S+(10% oil+5% water).

Here, the results of reconstruction agree very well with the expected values.

## 4. Experimental setup and results

The experiment was performed using the CV28 isochronous cyclotron at Physikalisch-Technische Bundesanstalt (PTB), Braunschweig, Germany. Neutrons were produced by a 12 MeV deuterium beam impinging on a 3 mm thick Be target. The beam was pulsed at a pulse repetition rate of 2 MHz frequency and pulse width of 1.7 ns. Average beam current was approximately 2 µA. The useful part of the neutron energy spectrum ranges from ca. 1 MeV up to 10 MeV (Brede et al, 1989). Neutron spectroscopy was performed by the time-of-flight (TOF) method using a cylindrical 2"x2" liquid scintillator detector (NE213 type) positioned at 11.5 m from the target.

Since we used a large non-pixelated detector, the samples were homogenously mixed and of constant thickness. They consisted of thin-walled, rectangular metal boxes 7.8x10.9x9.0 $cm^3$ in dimensions, filled with: dry sand (bulk density 1.55 g/cc), pure used motor oil (10W40, density 0.857 g/cc), pure water, or mixtures of dry sand containing 10.4%, 22.4% oil and 11.1% water by weight. The sample thickness traversed by the neutron beam was 7.8 cm. The distance between the sample and the detector was 65 cm. The acquisition time for the runs ranged from 1000 to 1200 s/run. The pure samples of dry sand, oil and water were used to determine the mass attenuation coefficients of each substance, which are required for reconstruction.

Fig. 7 shows an example of a TOF spectrum of flat (no sample), dry sand (S), a mixture of S+10.4% by weight motor oil, S+22.4% oil and S+11.1% water. The sharp peak on the left side of the spectra is the gamma-flash due to gamma-rays emitted from the target. The total number of detected neutrons (after subtracting the gamma-ray peak counts from the total) was $6 \cdot 10^7$ and $1.7-3 \cdot 10^7$ without and with samples respectively. As expected, the TOF spectra transmitted through the samples are dominated by the silica features. The spectra of S+10.4% oil and S+11.1% water appear to be quite similar in most parts of the spectrum, except for the peaks around channels 450 and 730. The latter correspond to carbon resonances at around 3 MeV.

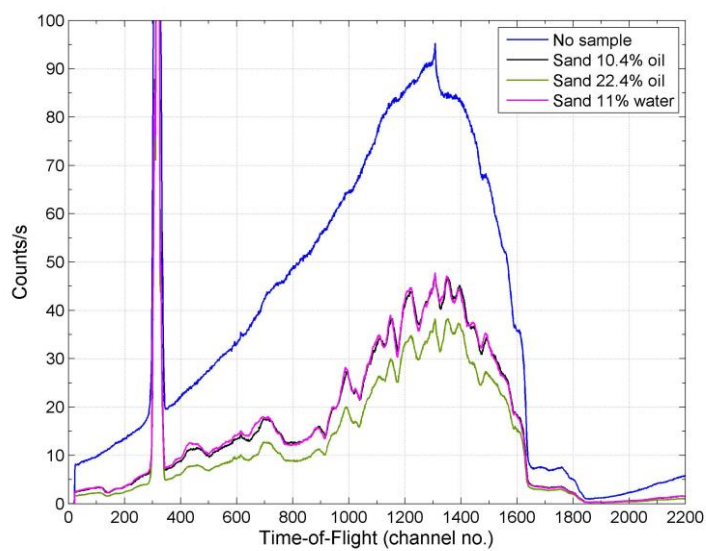

Fig. 7 Experimental neutron time-of flight spectra transmitted through: blue-flat spectrum (without sample), black: sand +10.4% motor oil. Purple: sand+11.1% water and green: sand+22.4% oil. The sharp peak on the left is the gamma-ray flash peak.

The transmission TOF spectra of dry sand, pure oil and water were used to determine the experimental mass attenuation coefficient of each substance and are shown in Fig 8.

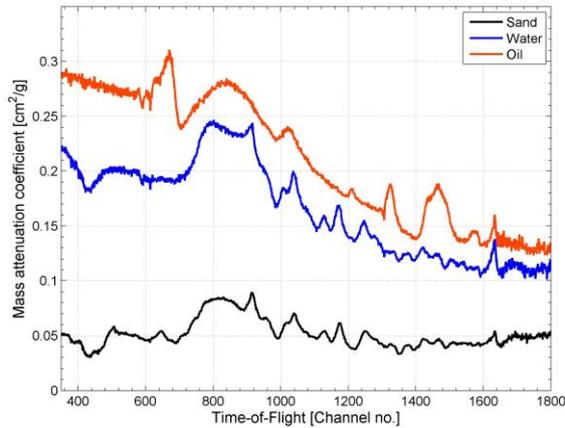

Fig. 8 Experimentally determined mass attenuation coefficients for dry sand, pure car engine oil and pure water.

Although the sharp resonances are less pronounced here than in the ENDF data of Fig. 1, the general features are similar.

Fig. 9 shows the reconstructed experimental areal density distributions of sand, oil and water in each sample. The distributions of the pure sand, sand+22.4% oil and sand+11.1% water follow quite well the expected behaviour (densities that peak at or around 0 and are skewed for the non-existent component), however the water distribution of the sand+10.4% oil sample shows an unexpected pattern, as if a significant amount of water is present in the sample, although the large error of ± 24% in this result compared to typically 5-8% uncertainty for an existing component may indicate that the reconstructed water density value might be consistent with zero, within statistical errors.

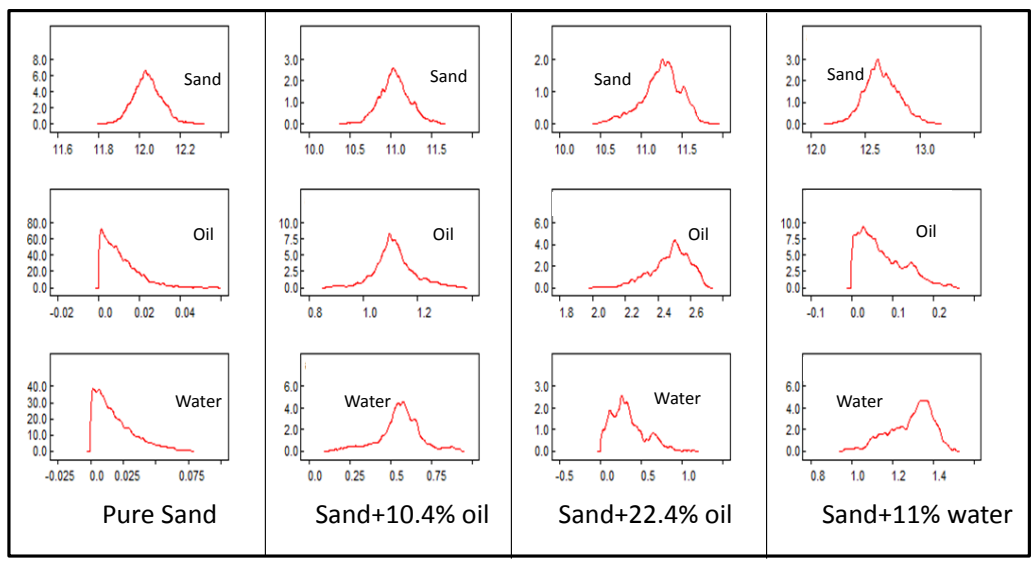

Fig. 9. Experimental areal density frequency distributions of sand, oil and water in: Pure sand, Sand+10.4% oil, sand +22.4% oil and sand+11% water. N=6·10$^7$ neutrons.

Table 3 summarizes the results obtained in the experiment and Fig 10 shows the empirical percentages by weight of oil and water in each configuration.

**Table 3:** Summary of reconstructed areal densities of sand samples

| Configuration | Sand [g/cm$^2$] | Oil [g/cm$^2$] | Water [g/cm$^2$] |
|---|---|---|---|
| **Pure Sand** | 12.04±0.07 (12.04*) | 0.01±0.008 (0*) | 0.018±0.015 (0*) |
| **S+10.4% oil** | 11.05±0.186 (12.04*) | 1.11±0.073 (1.3*) | 0.55±0.13 (0*) |
| **S+22.4% oil** | 11.24±0.24 (12.04*) | 2.48±0.12 (2.7*) | 0.33±0.21 (0*) |
| **S+11.1% water** | 12.65±0.157 (11.9*) | 0.073±0.055 (0*) | 1.29±0.11 (1.32*) |

*Expected value

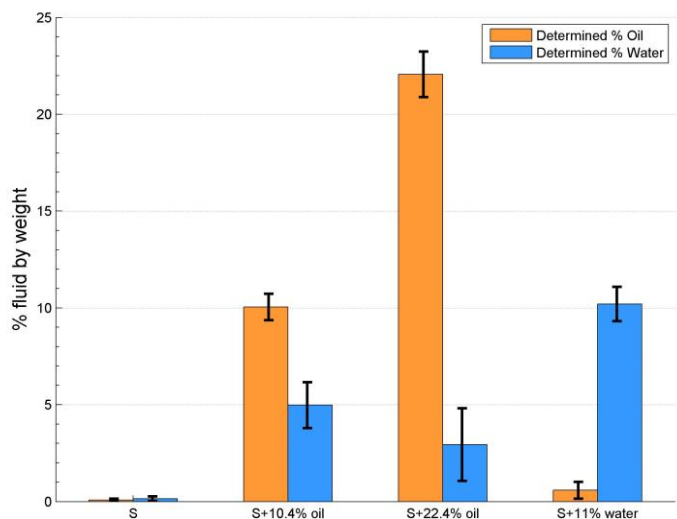

Fig. 10 Experimentally determined weight percentage of oil and water in sand samples.

## 5. Discussion and conclusions

We describe a method of Fast-Neutron Resonance Transmission (FNRT) radiography for a non-destructive, specific and quantitative determination of oil and water content in core samples and provide preliminary results of simulations and accelerator studies of laboratory samples.

Monte-Carlo calculations indicated that the fluid content can be determined with sufficiently high accuracy and precision by irradiating a 10 cm thick sandstone core with a broad energy neutron spectrum with a total of $10^6$ neutrons

Our experimental results generally confirm the Monte-Carlo calculations, but the standard deviations (precision of the measurement) of the reconstructed experimental values appear to be higher than those of the simulated results for similar incident neutron counts. This could be due to the different shape of the incident spectrum.

The case of sand with 10.4% oil shows unusually high (5%) content of water, albeit with an appreciably high relative uncertainty (±24%). The reason for this problem could be possible sample non-uniformity, due to imperfect mixing of the oil-sand mixture. The case of sand with 22.4% oil was already saturated and therefore well mixed. Water, being less viscous than oil mixes quite well with sand.

These effects and the influence of neutron scattering, background and detector pixel size need to be further investigated.

Provided the detector pixels are sufficiently small, the method measures the average liquid/dry-core-weight ratio in the path traversed by the fast neutrons regardless of the object shape, thickness or distribution. In principle the entire length of an intact core, within its protective sleeve can be scanned along the core length, providing information about the content distribution. The fluid weight fractions in the sample are determined independently, thus the ratio of oil-to-rock weight-ratio is independent of the water content. This is not the case in conventional thermal neutron radiography, where the assigned water and oil contents are unavoidably correlated due to their sensitivity to hydrogen content only. In addition, the use of fast neutrons can be useful in screening bulky objects such as thick rock cores, for which alternative probes, such as slow and epithermal neutrons, as well as low-energy X-rays, not only do not distinguish between hydrocarbons and water, but also suffer from limited penetration.

Traditional core analysis methods appear to be inadequate for tight shales which have very small porosity and permeability and new ways such as crushing the rock materials are performed in order to enable better access to the porous space. (Handwerger et al. 2012, Simpson and Fishman, 2015). In principle FNRT radiography method is suitable also for non-destructive fluid content determination in tight shales, however the determination of the small fluid content will require longer irradiation times for adequate statistical precision.

A prerequisite of the FNRT is a precise measurement of the incident and transmitted neutron spectra. We have applied time-of-flight (TOF) neutron spectroscopy, using a nanosecond-pulsed accelerator. In order to render an FNRT system more suitable for field applications, a pulsed T-T neutron generator, capable of providing a flat neutron spectrum up to 9 MeV or a tagged isotopic neutron source, such as $^{252}$Cf (Viesti et al, 2008) or $^{241}$Am-Be (Scherzinger J. et al, 2015) could be applied.